# Sheet Dependence on Superconducting Gap in Oxygen-Deficient Iron-based Oxypnictide Superconductors NdFeAsO$_{0.85}$

Yoshihiro AIURA[1,2], Koji SATO[1,3], Hideaki IWASAWA[2], Yosuke NAKASHIMA[4], Akihiro INO[4], Masashi ARITA[2], Kenya SHIMADA[2], Hirofumi NAMATAME[2], Masaki TANIGUCHI[2,4], Izumi HASE[1], Kiichi MIYAZAWA[1,5], Parasharam M. SHIRAGE[1], Hiroshi EISAKI[1], Hijiri KITO[1], Akira IYO[1,5]

[1] *National Institute of Advanced Industrial Science and Technology, Tsukuba, Ibaraki 305-8568*

[2] *Hiroshima Synchrotron Radiation Center, Hiroshima University, Higashi-Hiroshima 739-0046, Japan*

[3] *Faculty of Science, Ibaraki University, Mito, Ibaraki 310-8512, Japan*

[4] *Graduate School of Science, Hiroshima University, Higashi-Hiroshima 739-8526, Japan*

[5] *Department of Applied Electronics, Tokyo University of Science, 2641 Yamazaki, Noda, Chiba 278-3510, Japan*

## ABSTRACT

Photoemission spectroscopy with low-energy tunable photons on oxygen-deficient iron-based oxypnictide superconductors NdFeAsO$_{0.85}$ ($T_c$=52K) reveals a distinct photon-energy dependence of the electronic structure near the Fermi level ($E_F$). A clear shift of the leading-edge can be observed in the superconducting states with 9.5 eV photons, while a clear Fermi cutoff with little leading-edge shift can be observed with 6.0 eV photons. The results are indicative of the superconducting gap opening not on the hole-like ones around Γ (0,0) point but on the electron-like sheets around M (π,π) point.

Recent discovery of superconductivity in iron-based compounds brought about the frantic competition in condensed-matter physics [1], because this new superconductor is one of the most prospective materials with high temperature superconductivity except cuprate. Theoretical calculations showed two types of manifold Fermi surface (FS) sheets; three hole-like FS sheets around the Γ (0,0) point and two electron-like FS sheets around the M (π,π) points [2, 3]. To elucidate the role of each FS sheets for the superconducting gap may give a clue to elucidate the paring mechanism.

Although angle-integrated photoemission spectroscopy (PES) directly allows us to measure the electronic structure and the energy gap, the electronic contributions from those two types of FS sheets are merged in the final states when we used the conventional photon source like a helium discharge lamp. On the other hand, the low-energy tunable photons are very useful for separating out those contributions. With decreasing the photon energy extremely, the photoelectrons are to be observable only from the limited momentum space centered at the Γ point as shown in Fig. 1. Such low-energy high-resolution photoemission measurements have recently performed using a laser giving the photon energy of about 7 eV [4, 5]. These studies found a clear pseudogap but did not obtain any sign for the superconducting gap opening. Considering the probing area provided by 7 eV photons, there is no superconducting gap for the hole-like FS sheets around the Γ point. It is thus essential to elucidate the characteristic electronic structure due to the electron-like FS sheets around the M point, in order to understand the pairing mechanism and origin of the superconductivity.

In this letter, we report high-resolution photoemission spectroscopy with low-energy tunable photons (LEPES) on oxygen-deficient iron-based oxypnictide superconductors $NdFeAsO_{0.85}$. We have found a clear shift of the leading-edge away from $E_F$ in the superconducting state using 9.5 eV photons. On the other hand, a clear Fermi cutoff with little leading-edge shift can be observed in the case of 6 eV, which is qualitatively consistent with previous laser based photoemission results [4, 5]. Present LEPES results demonstrate that the superconducting gap opens on the electron-like FS sheets around the M point, but not on the hole-like ones around the Γ point, suggesting that different type of Fermi surface sheets may have a different role in the superconducting states.

Polycrystalline samples of $NdFeAsO_{1-\delta}$ were synthesized at high pressure and high temperature using a cubic-anvil high-pressure apparatus. Details of synthesis method are given in [6]. Although the nominal oxygen deficiency in the $NdFeAsO_{1-\delta}$ samples is 0.35 estimated from the starting materials, recent powder neutron diffraction revealed that the samples are largely shift towards higher oxidation [7].

Judging from the relationship between the lattice parameters and the oxygen deficient content, the actual oxygen content used in this experiment is estimated to be 0.85. The sharp drop of the magnetic susceptibility, corresponding to the onset of superconductivity, is observed at 52 K. We note that the amount of induced carriers by the oxygen deficiency should be two times larger than that by the oxygen substitution in fluorine-doped NdFeAsO$_{1-x}$F$_x$ [8], assuming that these dopants such as oxygen deficiency or fluorine provide the carriers (electrons) expected from the formal valance into the conduction layer, FeAs. LEPES measurements were performed at beamline 9A of Hiroshima Synchrotron Radiation Center (HiSOR) using a Scienta R4000 electron analyzer. The sample was fractured *in–situ* and then measured below ultra-high vacuum (~3×10$^{-11}$mbar).

Figure 1 shows FSs of LaFeAsO obtained by projecting all $k_Z$ into the $k_Z$=0 plane, where the probing area at low-photon energies (6-10 eV) are indicated by the circles. As shown in this figure, it is expected that the probing area covers the entire first Brillouin zone above 9 eV, and the 7 eV is the boundary energy whether the electron-like FSs around the M point to be detected or not.

Figure 2 (a) shows the hv-dependent near-E$_F$ LEPES spectra from NdFeAsO$_{0.85}$ taken at 10 K, where all the spectra are normalized by the photon flux. At first sight, one can see a clear Fermi cutoff in the case of 6 eV, indicating the crossing E$_F$ of the hole-like FSs centered at Γ. This is qualitatively consistent with our simple estimation based on a rigid band model, where the hole-like FS sheets are drastically shrinking with doping the electrons but survive even for 0.3 electrons in NdFeAsO$_{0.85}$. According to previous PES studies [9, 10], the near-E$_F$ electronic structure consists almost exclusively of the Fe 3d states. On the other hand, we found that the near-E$_F$ spectral weight is drastically increasing with decreasing the photon energy in opposite to the behavior of the photoionization cross sections of the Fe 3d states. This variation would be thus caused by the final state effects as shown in cuprates [11, 12] or the As 4p contributions for the near-E$_F$ electronic structure, which becomes significantly important for low photon energy regime below 25 eV [9]. To identify the contribution due to the As 4p states, we measured the hv dependence of LEPES spectra up to 30 eV. Then we found the large increase in the spectral weight around the photon energy of 12 eV as shown in Fig. 2 (b), which may be attributed to the As 4s-4p resonance enhancement [13]. These results suggests the existence of the As 4p contribution hybridized with the Fe 3d sates in the vicinity of E$_F$.

To remove the thermal broadening effect and investigate the variation of near-E$_F$ spectral features in detail, we symmetrized the LEPES spectra with respect to

$E_F$ using hv=6 eV (red curve) and 10 eV (black curve) as shown in Fig.2 (c). The symmetrized spectra have been normalized by the integrated intensity between -0.1 eV and 0.1eV to compare the detailed variation near $E_F$ clearly. We found two types of spectral lineshape from symmetrized spectra; a "V"-shaped lineshape can be observed with hv=6 eV but the gap-like spectra feature near $E_F$ is clearly shown with hv= 10eV. Roughly speaking, ~ 7 eV seems to be boundary again between two lineshapes like the probing area as mentioned above. Namely, the observed density of states above 7 eV is not solely due to the hole-like FS sheets centered at the Γ point but also due to the electron-like FS sheets around the M point. In contrast, the hole-like FS sheets centered at the Γ point mainly contributes to the electronic states in the case of below ~7 eV. All of these are implying that the electronic state due to the electron-like FS sheets around the M point is an important key to understand the superconducting gap and pseudogap.

Figure 3 (a) shows the temperature-dependence of LEPES spectra of NdFeAsO$_{0.85}$ using 9.5 eV photons, which allow the photoelectrons to be emitted from the entire first Brillouin zone (Fig. 1). Surprisingly, the crossing point of LEPES spectra is significantly deep in energy far below $E_F$. As indicated by an arrow in the inset to the figure, the crossing point is located at ~ 4 meV, which is 10 times larger than previously reported one in LaFeAsO$_{0.93}$F$_{0.07}$ with T$_c$=24 [14]. In addition, it is shown that the spectral intensity at $E_F$ systematically recovers with increasing temperature. These results indicate a clear evolution in the opening of the superconducting gap as a function of temperature. Although the similar variation in the spectral intensity near the leading edge is shown in conventional superconductors [15], no clear superconducting coherence peak is observed in LEPES spectra of iron-based layered superconductor Nd Fe AsO$_{0.85}$, which may be caused by disorder effects in this system as shown in boron-doped diamond [16].

On the contrary, recent ARPES studies of Ba$_{0.6}$K$_{0.4}$Fe$_2$As$_2$ showed multiple nodeless superconducting gaps with clear coherence peak below T$_c$ [17, 18]. The lack of the superconducting coherence peak in our LEPES spectra may be derived from the existence of pseudogap, which is suggested by a clear spectral intensity at $E_F$ even well below T$_c$ as shown in Fig. 3 (a). The existence of the pseudogap is already confirmed via previous PES studies [4, 5, 14, 17, 19, 20]. To further examine the opening of the superconducting gap, LEPES spectra are symmetrized with respect to $E_F$, as shown in Fig.3 (b). Again one can see the clear depletion of spectral weight in the vicinity of $E_F$ with decreasing temperature.

Turning now to LEPES spectra from 6 eV photons shown in Fig. 3 (c), a clear Fermi cutoff with little leading-edge shift can be observed, which is qualitatively consistent with previous laser based photoemission results [4, 5]. Symmetrized spectra from the spectra hν=6 eV with respect to $E_F$ is shown in Fig. 3 (d). The suppression in the spectral intensity near $E_F$ is observed with decreasing temperature again, but is rather smaller than that with hν=9.5 eV photons. The near-$E_F$ LEPES spectra with 6 eV photons suggest the existence of a pseudogap but no superconducting gap on the hole-like FS sheets around Γ point.

In conclusion, we performed high-resolution photoemission spectroscopy with low-energy tunable photons on oxygen-deficient iron-based oxypnictide superconductors NdFeAsO$_{0.85}$. We found a clear leading-edge shift far below $E_F$ in the superconducting state using 9.5 eV photons, but not found in the case of 6 eV photons. This suggests that the superconducting gap opens on the electron-like FS sheets around the M point, but not on the hole-like ones around Γ point.

## Acknowledgement

This work was supported by Research Fellowships of the Japan Society for the Promotion of Science for Young Scientists and Grant-in-Aid for COE research (No.13CE2002) of MEXT Japan. The synchrotron radiation experiments have been done under the approval of HSRC (Proposal No. 08-A-4).

Figure caption

Fig. 1   Fermi surface for all $k_z$ projected on to the $k_z$=0 plane. The circles mean the probing area of LEPES spectra for the photon energy from 6 eV to 10 eV.

Fig. 2   (a) hν dependence of LEPES spectra near $E_F$ of NdFeAsO$_{0.85}$ taken at 10 K. The spectra normalized by photon flux are shown from hν=6 eV to 10 eV every 0.2 eV step. The energy resolution was set to 8 meV. (b) Integrated intensity of LEPES spectra from $E_F$ to -0.05 eV as functions of photon energy. (c) Symmetrized spectra with respect with $E_F$ using hν=6eV (red curve) and 10 eV (black curve). The spectra are normalized by the integrated intensity from -0.1 eV to 0.1 eV.

Fig. 3   (a) Temperature dependence of photoemission spectra of NdFeAsO$_{0.85}$ ($T_C$=52 K) taken at hν=9.5 eV. Expanded spectra in the vicinity of $E_F$ are shown in the inset. The energy resolution was set to 5 meV. (b) Symmetrized spectra from the spectra hν=9.5eV in (a) at with respect with $E_F$. (c) and (d) Same as in (a) and (b), respectively, except the photon energy is now changed to 6 eV. The overall energy resolution was set to 3 meV.

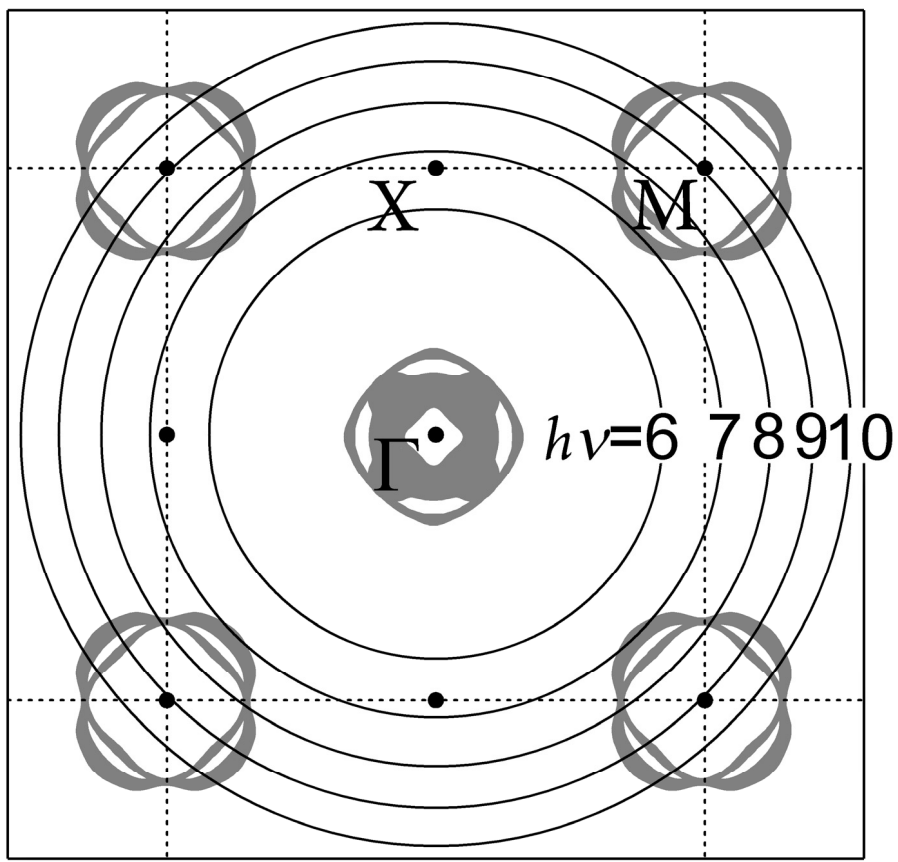

Fig. 1　Aiura *et al.*

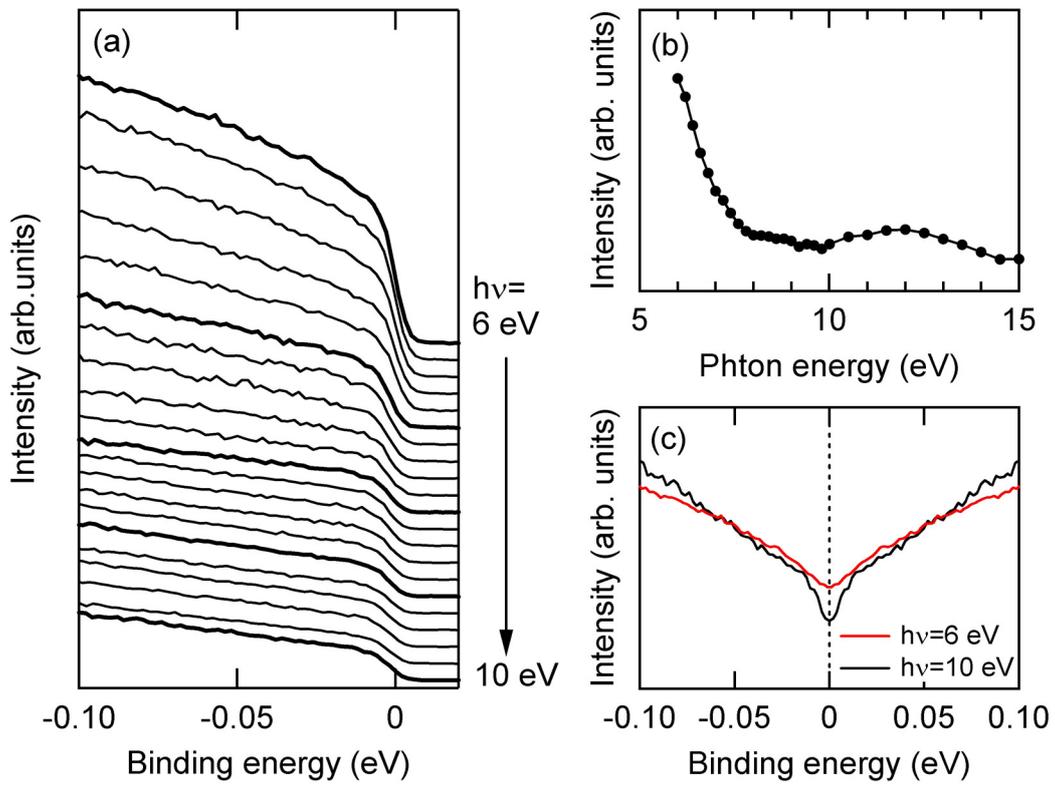

Fig. 2    Aiura *et al.*

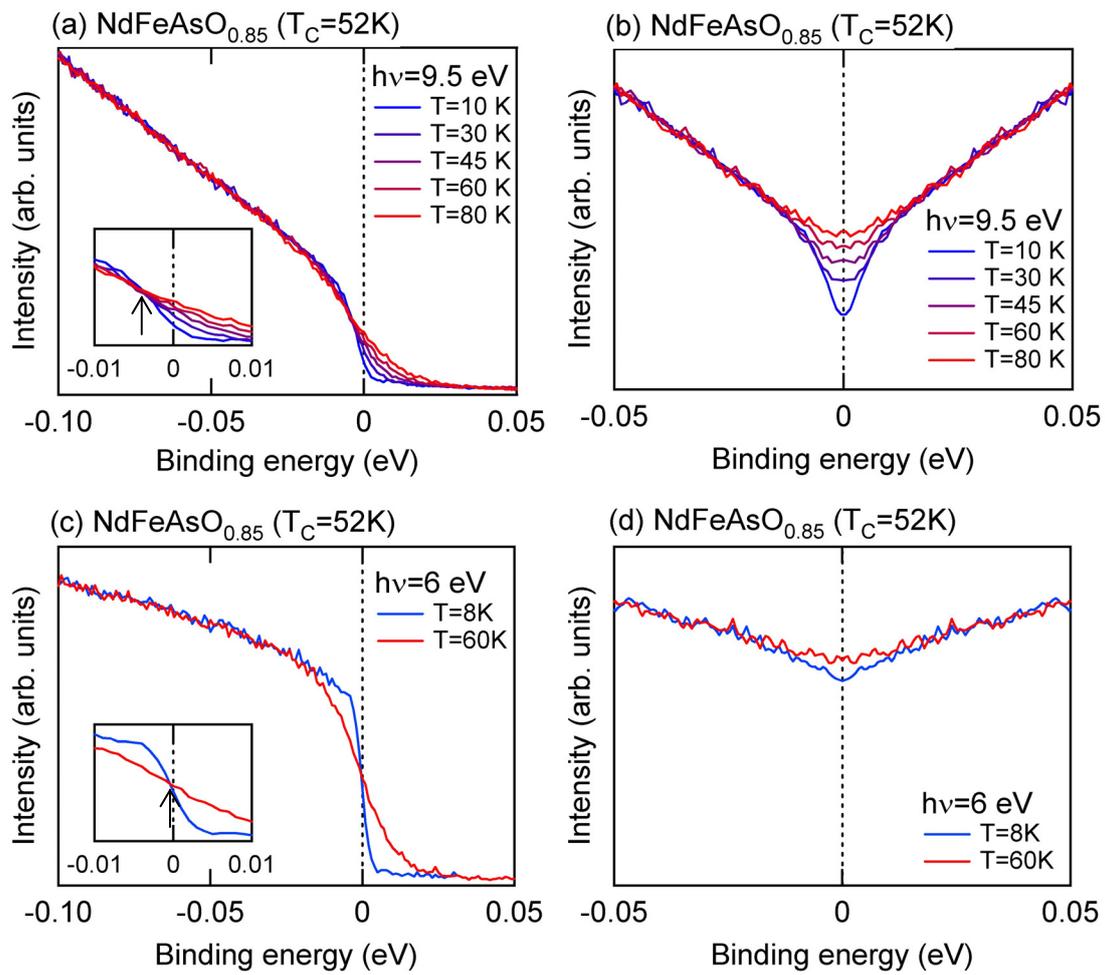

Fig. 3    Aiura *et al.*